\documentclass[aps,nofootinbib,showpacs,showkeys,twocolumn,tightenlines,preprintnumbers]
{revtex4}

\usepackage{epsf,epsfig,subfigure,axodraw,graphicx,amsmath,amssymb}

\begin{document}
\preprint{SNUTP 06-004}
\title{\Large\bf Modification of Decay Constants of Superstring
Axions: \\ Effects of Flux Compactification and Axion Mixing}
\author{ Ian-Woo Kim\footnote{iwkim@phya.snu.ac.kr} and
Jihn E.  Kim\footnote{jekim@phyp.snu.ac.kr} }
\address{ School of
Physics and Astronomy, and Center for Theoretical Physics, Seoul
National University, Seoul 151-747, Korea }


\begin{abstract}
We study possibilities for lowering the decay constants of
superstring axions. In the heterotic Calabi-Yau compactification, a
localized model-dependent axion can appear at a nearly collapsing
2-cycle. The effect of flux can be used for generating warp factor
suppression of the axion decay constant. We also point out that the
hidden sector instanton potential much higher than the QCD instanton
potential picks up the larger effective axion decay constant as that
of the QCD axion. We show that this can be converted by introducing
many hidden-sector quarks so that the decay constant of the QCD
axion turns out to be much smaller than the string scale.
\end{abstract}

 \maketitle


\def\lsl{ l \hspace{-0.45 em}/}
\def\ksl{ k \hspace{-0.45 em}/}
\def\qsl{ q \hspace{-0.45 em}/}
\def\psl{ p \hspace{-0.45 em}/}
\def\ppsl{ p' \hspace{-0.70 em}/}
\def\dsl{ \partial \hspace{-0.5 em}/}
\def\Dsl{ D \hspace{-0.55 em}/}
\def\N{$\cal N$}
\def\tphi{\tilde\phi}

\def\f#1#2{\frac{#1}{#2}}


\section{introduction}

One of the most puzzling problems of the standard model is the
strong CP problem. There are a few appealing solutions on the strong
CP problem \cite{kimprp87}, among which the most attractive and
promising one is considered to be  a spontaneously broken
Peccei-Quinn(PQ) symmetry \cite{peccei}. At present, the resulting
invisible axion is phenomenologically allowed only within the window
$10^{9}\ {\rm GeV}\le F_a\le 10^{12}\ {\rm GeV}$, with the upper end
plausible for the axion to be a candidate of dark matter. If so, it
is better for the invisible axion to be realized in the most
fundamental theory such as in superstring theory. Indeed,
superstring theories provide such a candidate in the components of
the antisymmetric tensor field $B_{MN}\ (M,N=0,1,\cdots,9)$, in
terms of the model-independent axion(MI-axion) \cite{Witten84}
and/or the model-dependent axions(MD-axion) \cite{Witten85}.

When we consider unbroken nonabelian groups, there appear
corresponding $\theta$ parameters and as such the same number of
axions is required to settle all the $\theta$ parameters. Our
interest regarding the strong CP problem is the QCD axion for which
the afore-mentioned window for the axion decay constant is
applicable. For the MI-axion, the decay constant is of order
$10^{16}$~GeV which is harmful if it is interpreted as the QCD axion
\cite{kim85}. For the MD-axions, the decay constants take generic
values of order the string scale since it is the only scale of the
theory. If there exist several axions, axion masses are determined
by the scales of the corresponding nonabelian gauge groups and also
by the axion decay constants. If some plausible cancellation occurs
such that one axion is almost a Goldstone boson \cite{nilles03}, the
decay constant of the corresponding axion can be much bigger than
the Planck scale, which is desirable for it to be a quintessential
axion. For some inflationary models, such a large decay constant is
needed \cite{nilles03}.  But generically, we expect that the decay
constants are of order the string scale \cite{kim85}.

Therefore, it is desirable to have some smaller scales for decay
constants of superstring axions. Recently, generating such
hierarchical scales has been explored in the extra-dimensional
context. Those mechanisms involve the localization of dynamical
degrees of freedom at some location in compactified space.
Especially, the localization of the gravity enables us to scale down
the overall scales of physics of interest, introducing a warp factor
\cite{RSI}.

The well-known warping from string models arises from the
Klebanov-Strassler(KS) throat geometry \cite{KS98}. In Type IIB
string models, Giddings, Kachru and Polchinski(GKP) localized the
$\langle F_{ijk}\rangle$ flux by the balancing act of the $\langle
H_{ijk}\rangle$ flux \cite{GKP02} so that stringy realization of the
Randall-Sundrum type model with stabilization mechanism has been
successfully constructed. In heterotic strings, supersymmetric
compactification with flux yields a warped product of 4D Minkowski
spacetime and non-K\"ahler internal manifold, where the necessary
conditions for a warp factor were given a long time ago by
Strominger \cite{Strominger86}. Recently Becker {\it et al.} worked
on compactification of heterotic strings on the non-K\"ahler
manifold with the use of fluxes of antisymmetric tensor field
$B_{MN}$ \cite{Becker}. Here, the interpretation of the warp factor
in the closed string theory assumes some kind of localization of
wave functions.

Our prime objective is to find out mechanisms of lowering axion
decay constants. One easy method is an intermediate scale
compactification \cite{intst}. In this paper, we look for {\it a
possibility of warping the effective axion decay constants} in
compactifications of string models. Previous discussions
\cite{kim85,choikim85,kim99,choiM} on superstring axions did not
consider the effects of warp factors and the localization of
superstring axions. If the background is unwarped, it is unavoidable
to have the axion decay constant at the string scale unless we deal
with a large volume compactification. To have a large hierarchy of
axion decay constants, we emphasize that a warped background is
effective for some localized MD-axions.

In the end, we discuss the effect of axion mixing. We  point out
that the hidden sector instanton potential much higher than the QCD
instanton potential leads to the larger effective axion decay
constant as that of the QCD axion due to the axion mixing. However,
it is possible to choose the smaller decay constant as the QCD axion
decay constant by introducing many light hidden-sector quarks.

\section{Warping in heterotic string models}

In this section, we consider the heterotic string theory
compactified with flux.\footnote{Recently, there has been a detailed
analysis on lowering axion decay constant in type II string theories
by Conlon \cite{conlon}.} In compactification of string models, it
is required for the vacuum to possess a four dimensional(4D) \N=1
supersymmetry after compactification, which is studied from the
$D=10$ supersymmetry transformation of Fermi fields: gravitino
$\psi_M$, dilatino $\lambda$ and gaugino $\xi$ which transform by
\begin{align}
\begin{split}
&\delta_\epsilon \psi_M=\textstyle \nabla_M \epsilon+\frac14 {\bf
H}_M
 \epsilon,\\
&\delta_\epsilon \lambda=\textstyle  (\dsl\phi) \epsilon+\frac12
 {\bf H}\epsilon,\\
&\delta_\epsilon \xi=2{\bf F}\epsilon,
\end{split}
\end{align}
where $\phi$ is dilaton, $F_{MN}$ is the field strengths of gauge
fields and 3-form flux $H_{MNP}$ is defined in terms of NS 2-form
potential $B_{MN}$ and the Yang-Mills Chern-Simons term $\omega_Y$
and Lorentz Chern-Simons term $\omega_L$ according to
\begin{equation}
H = dB + \frac{\alpha'}{4} \left( \frac{1}{30} \omega_Y
 - \omega_L \right)
\end{equation}
Here, the bold faced ones are the
field strengths contracted with appropriate 10D $\gamma$ matrices,
for example ${\bf H}_M=H_{MNP}\gamma^{NP}$.

The pioneering work of Candelas {\it et al.} \cite{CHSW} searched
for vacua having non-zero covariantly constant supersymmetry
parameter $\epsilon$ in case fluxes vanish, ${\bf H}_M=0$ and ${\bf
H}=0$, and obtained the background manifold with K\"ahler geometry.
For the case of 6D internal space, it is the Calabi-Yau space which
has $SU(3)$ holonomy.

More general conditions required for 4D \N=1 supersymmetry with
non-vanishing fluxes have been obtained in Ref. \cite{Strominger86}.
The recent work \cite{Becker} obtained non-singular solutions
satisfying Strominger's four conditions. With non-vanishing fluxes,
the internal manifold turns out to be non-K\"ahler.

The geometry of compactification and the dynamics of low energy fields
in the string theory are described by 10D low energy supergravity
whose bosonic part is
\begin{eqnarray}
&&S= \frac{1}{2\kappa_{10}^2}\int d^{10}x\sqrt{-g}e^{-2\phi}
\times \nonumber \\
&&\quad \left({\cal{R}} +4|\partial\phi|^2-\frac12 |H|^2
+\frac{\alpha^\prime}{4}
 {\bf tr}|F|^2 +2{\cal L}_{\rm GS}\right)\label{action}
\end{eqnarray}
where $\kappa_{10}$ is the inverse of reduced 10D Planck mass, and
 ${\cal L}_{\rm GS}$ is the Green-Schwarz term.

The MI-axion is the antisymmetric tensor field in the tangent 4D
space. Its equation of motion is determined universally from the
Bianchi identity:
\begin{equation}
dH=\textstyle R\wedge R-\frac{1}{30}F\wedge F,
\end{equation}
i.e. the $H_{\mu\nu\rho}$ coupling to the gluon field strength is
\begin{equation}
\textstyle\frac{1}{3!}\epsilon^{\mu\nu\rho\sigma}\partial_\mu
H_{\nu\rho\sigma} =\textstyle -\frac{1}{30}(\frac{1}{2!})^2
\epsilon^{\mu\nu\rho\sigma}F_{\mu\nu}F_{\rho\sigma}. \label{MIeq}
\end{equation}
Defining $H_{\mu\nu\rho}=M
\epsilon_{\mu\nu\rho\sigma}\partial^\sigma a_{MI}$ in terms of the
properly normalized MI-axion $a_{MI}$, basically $M$ turns out to be
proportional to the MI-axion decay constant.
$M$ is determined by 4D Planck scale and the scale of the coefficient
of $|H|^2$ term in 4D effective action. From the ratio of warp
factors of the Einstein-Hilbert term and the $|H|^2$ term
in (\ref{action}), we find that the MI-axion does not have a warp
factor dependent decay constant. This is because the RHS of Eq.
(\ref{MIeq}) is a topological term.

For MD-axions, $B_{ij}$,  $i,j$ are internal space indices. For
these MD-axions,  there can be warp factor dependence. The MD-axions
couple to field strengths via the Green-Schwarz term \cite{GS84},
\begin{equation}
\int d^{10}x{\cal L}_{\rm GS}=c\int (-3BX_8+2X_3^0 X^0_7)
\end{equation}
for which the couplings of MD-axions are schematically given as
 \cite{GS84,Witten85,choikim85},
\begin{align}
\int B\wedge F\wedge F\wedge F\wedge F+\cdots.
\end{align}
Depending on the way $B_{ij}$ is embedded in the internal space,
MD-axions can be localized. The Hodge-Betti number $b_{(1,1)}$
corresponds to the number of MD-axions. In CY spaces, it is known
that $b_{(1,1)}\ge 1$ \cite{hubsch}; thus there exists at least one
MD-axion. The MD-axion in the simplest case of $b_{(1,1)}=1$
corresponds to the breathing mode of the internal space. Even with
fluxes, this breathing mode couples to the E$_8$ and E$_8^\prime$
anomalies without exponentially small decay constant since the
breathing mode is not localized.

Recently, Becker {\it et al.} obtained an interesting non-K\"ahler
geometry with heterotic flux compactification \cite{Becker}. In the
string frame, the background is a direct product of 4D spacetime and
an internal six-dimensional manifold with the 10D metric
\begin{equation}
ds^2= \eta_{\mu\nu}dx^\mu dx^\nu+ds_{(6)}^2.
\end{equation}
The internal space geometry is a $T^2$ bundle over the base K3
manifold
\begin{equation}
ds_{(6)}^2=e^{2\phi}ds^2_{\rm
K3}+(dy_5+\alpha_5)^2+(dy_6+\alpha_6)^2 \label{base4D}
\end{equation}
where $ds_{\rm K3}^2$ is the metric for the four dimensional base
manifold ${\rm K3}$, with coordinates $y_1,\cdots,y_4$.  Here,
$\phi$ is dilaton and 1-forms $\alpha_5$ and $\alpha_6$ depend only
on the $K3$ coordinates. To obtain 4D \N = 1  SUSY, the torus must be
fibered nontrivially, which leads to the complication of the
solution. However, such twisting has only global effects.

To discuss the property of warped geometry arising near 2-cycles of
the above solution, it is enough to consider the trivially fibered
case. In the orbifold limit of $K3$-manifold \cite{Becker03},
$ds^2_{\rm K3}$ describes a flat 4-torus $T^4$ and
\begin{align}
\begin{split}
&\Delta^2 = c_0 + A z^1 + B z^2 + {\rm c.c.},
\quad\quad{\rm with}\  \Delta=e^\phi, \\
& H= A dz^1 \wedge dz^2 \wedge d\bar{z}^2
 + B dz^1 \wedge d\bar{z}^1 \wedge dz^2 + {\rm c.c.}
 \end{split}
\end{align}
where $z^1$ and $z^2$ are the coordinate of $T^4$. The 10D metric in
the Einstein frame is given by
\begin{align}
&ds^2_E = \Delta^{-1/2} (\eta_{\mu\nu}dx^\mu dx^\nu)\\
& + \Delta^{-1/2} \left( (dx + \alpha_1)^2 + (dy+\alpha_2)^2 \right)
+ \Delta^{3/2} ds^2_{\rm K3}.\nonumber
\end{align}
One can easily see that the Becker {\it et al.} solution gives a
power-law warp factor  for $c_0 \ll A z^1 + B z^2 $. Even though we
have not obtained an exponential warp factor here, exponential warp
factors may arise in some other flux compactifications.

\begin{figure}[t]
\centering \epsfig{figure=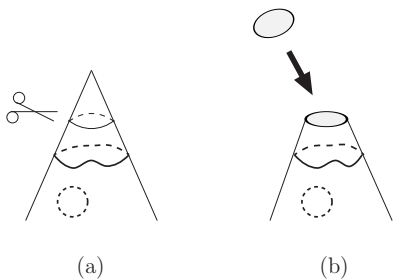, width=6cm,
bbllx=180,bblly=620, bburx=420,bbury=770 } \caption{The Calabi-Yau
twofold is obtained by blowing up $T^4/Z_2$ orbifold. This is
achieved by cutting around a singular fixed point and replacing it
with a smooth manifold (squeezed throat), which is also moded out by
the same $Z_2$.}\label{Blowingup}
\end{figure}

The warp factor effect is significant for localized fields
\cite{RSI}.  Strings on fixed points (in the twisted sector) in the
orbifold compactification are considered to be localized, and hence
can have warp factors if the orbifold is a limit of some manifold
with fluxes. These localized strings cannot be moved to the bulk
because of the closed string condition, satisfying the modding
group, around the fixed point.  For some CY spaces, there exist the
corresponding orbifolds from which the  CY space can be made by
blowing up the fixed points as shown for $T^4/Z_2$ orbifold in Fig.
\ref{Blowingup}. Let us call the corresponding region in the CY
space `squeezed throat'. Around the fixed point of the orbifold and
hence around the cut line of the corresponding CY space, topological
invariants must be the same. Namely, the curve around the fixed
point (solid and dashed line ring of Fig. \ref{Blowingup}(a)) and
the closed ring in the bulk (the dashed line ring of  Fig.
\ref{Blowingup}(a)) have different topological properties and hence
from the topological argument the solid-dash ring cannot be moved to
the dashed ring without obstruction of topology. Even though there
is no singularity in the corresponding CY space as shown in Fig.
\ref{Blowingup}(b), the solid-dash ring of Fig. \ref{Blowingup}(b)
cannot be moved to the dashed ring of Fig. \ref{Blowingup}(b)
without obstruction of topology, because the topological number must
change  in the CY space also for this to happen.

Now from physical argument, we can guess which MD-axions can have
warp factors. The MD-axion corresponding to the breathing mode must
belong to the cycle encompassing the whole internal space, and
hence it does not belong to a localized string. So the breathing
mode MD-axion would not have a warp factor. The remaining MD-axions
belonging to some squeezed throat of the CY space and can have warp
factors. The maximum number of MD-axions having the warp factors is
\begin{equation}
b_{(1,1)}-1.
\end{equation}
Even if a MD-axion is localized at a fixed point, it does not
necessarily mean that it has a warp factor. The additional condition
is to have a stable squeezed throat {\it \`a la} KS and GKP.

To set up the dynamical scale of axion fields by warp factor, it is
crucial to understand how the wavefunction of a 4D massless field is
localized in the compactified internal manifold $X$. The zero modes
are determined by harmonic $p$-forms $\omega$ satisfying $\Delta_d
\omega = 0$ where the Laplacian
is defined by $\Delta_d = d^\dagger d + d d^\dagger =
(d+d^\dagger)^2$.
Here, the adjoint exterior derivative operator on $p$-form
 $\omega = \frac{1}{p!}
\omega_{{\mu_1}\dots{\mu_p}} dx^{\mu_1} \wedge \dots
 \wedge dx^{\mu_p} $
 is
\begin{equation}
d^\dagger \omega = \frac{1}{p!} \nabla^\mu \omega_{\mu {\nu_2}
\dots {\nu_p}} dx^{\nu_2} \wedge \dots \wedge
dx^{\nu_p}.
\end{equation}
The Laplacian acted on the $p$-form $\omega$ is written explicitly by
\begin{align}
\Delta_d \omega = \frac{1}{p!} &\Big( \nabla_{[ \mu_1 } \nabla^\nu
\omega_{\nu \mu_2 \dots \mu_p ] }\nonumber\\
& + \nabla^\nu \nabla_{[ \nu} \omega_{\mu_1 \dots \mu_p ] } \Big)
dx^{\mu_1} \wedge \dots \wedge dx^{\mu_p}.
\end{align}

Harmonic $p$-forms belong to the cohomology group $H^p(X)$ on $X$.
If we have
a set of $p$-cycles $\{ C_i\}$ on $X$, we find the basis
$\{\omega_j\}$ for the harmonic $p$-forms such that the period
satisfies
\begin{equation}
\int_{C_i} \omega_j = \delta_{ij}
\end{equation}
due to  de Rham's theorem. The harmonic $p$-forms are the zero mode
wave functions. The period integrated over the same homology class
where the integration is performed has the same value. This fact
indicates that the wave function has the largest value on the
smallest cycle in the same homology class. If a cycle can be
deformed to a nearly vanishing size, the zero mode wave function is
localized like a delta-function at that point, which is the case for
the twisted mode in the orbifold compactification.

The zero mode wave function of $E_8 \times E_8^{\prime}$ gauge
fields are determined by harmonic zero forms which cannot have the
localization effect discussed in the preceding paragraph. However,
the zero mode wave function of MD-axions are from harmonic 2-forms,
and thus it is important to examine the 2-cycles which can shrink to
a point where the effect of the warp factor is significantly large.

In summary, some MD-axion(s) in the heterotic string can be
localized and its scale can have a warp factor suppression. But, for
MD-axions, there may appear dangerously large superpotential terms
as pointed out by Wen and Witten \cite{WenWitten}. However, the size
of these terms depend on details of compactification.


\section{Axion mixing}

If there exist more than one axion, one crucial question is what is
the QCD axion and its decay constant. Even if the scale of one
MD-axion is warped to give an intermediate scale, it is not
automatic that the scale of the QCD axion is at the intermediate
scale because the scale of MI-axion is of order 10$^{16}$ GeV. A
naive guess leads to a 10$^{16}$ GeV for the scale of the QCD axion.

In fact, the above pessimistic comment is related to our another
method of lowering the axion decay constant.

Let us consider two nonabelian gauge groups, the hidden sector
group, say $SU(N)_h$, confining at the intermediate scale and QCD
$SU(3)_c$, and two axions $a_1$ and $a_2$ with decay constants $F_1$
and $F_2$ with a hierarchy $F_1\gg F_2$. Considering two axions is
equivalent to introducing two global symmetries. If mass eigenstates
are nontrivial mixtures of $a_1$ and $a_2$, then the higher
instanton potential corresponds to  the lower decay constant and the
lower instanton potential corresponds to the higher decay constant
\cite{kim99}.\footnote{In Ref. \cite{nilles03} the power was not
correctly given.} On the other hand, remember that if there exists a
massless quark then the instanton potential vanishes, which
corresponds to a flat axion potential. Therefore, for a light quark
mass $m_q$ the instanton potential is proportional to $m_q$
\cite{tHooft}. For $n$ light quarks, one might naively expect a
power law of $m_q^n$, which however is not warranted. This is
because there is only one $\theta$ parameter for a nonabelian group
and only one current out of $n$ quark currents contributes to the
axion potential.\footnote{The instanton potential for $\eta^\prime$
does not have a quark mass.} This is the reason that the QCD axion
potential has only one light quark mass factor \cite{kimprp87}. Here
also, this property of one power of current mass is not changed.

Naively, it is guessed that the hidden sector instanton potential is
much higher than the QCD instanton potential in which case the QCD
axion corresponds to the larger axion decay constant $F_1$, which is
harmful. But, if almost massless hidden-sector quarks exist, then
one can obtain a much lower hidden-sector instanton potential but
the suppression is only with one mass factor of hidden-sector
quarks.
But the following study shows that one must introduce sufficiently
many hidden-sector quarks.

\subsection{Analogy to old $U(1)$ problem, instanton potential
and $\eta^\prime$-like particle masses}

Our objective is to lower the hidden-sector instanton potential
below that of the QCD instanton potential. In analogy with the
resolution \cite{U1thooft} of the old $U(1)$ problem, we determine
the height of the hidden-sector axion potential. The relevant
Lagrangian with the hidden-sector SUSY $SU(N)_h$ gauge group with
one hidden-sector quark $q$ is
\begin{equation}
{\cal L}_\chi=-M\lambda\lambda -m\bar qq+{\cal L}_{\rm inst}+{\rm
h.c.} \label{IntL}
\end{equation}
where $\lambda$ is the hidden-sector gaugino, $q$ is the
hidden-sector quark, and ${\cal L}_{\rm inst}$ is the 't Hooft
determinental interaction,
\begin{equation}
{\cal L}_{\rm inst}
=e^{-\frac{8\pi^2}{g^2}-i\theta}(\lambda\lambda)^N(\bar qq)
\label{instInt}
\end{equation}
which shows that a chiral rotation by angle $\alpha$ on the
hidden-sector quark is equivalent to changing the $\theta$ parameter
by $\alpha$, and a chiral rotation by angle $\alpha$ on the
hidden-sector gaugino is equivalent to changing the $\theta$
parameter by $N\alpha$. The resolution of the old $U(1)$ problem
comes from the instanton term, which will be manifest in the
following discussion also.

\begin{figure}[t]
\centering \epsfig{figure=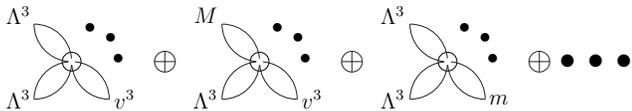, width=6cm,
bbllx=180,bblly=690, bburx=420,bbury=770 } \caption{The fermions of
the determinental interaction are closed by condensations or by
current masses. Condensations accompany phases which are not shown
explicitly.}\label{tHooft}
\end{figure}

The condensations of hidden-sector gluino and hidden-sector quark
produce composite pseudoscalar particles which are denoted as
$\eta_\lambda\equiv F_\lambda\theta_\lambda$ and $\eta_q\equiv
F_q\theta_q$, respectively, where $\langle\lambda\lambda\rangle
\simeq -\Lambda^3e^{i\theta_\lambda}$ and $\langle\bar qq\rangle
\simeq -v^3e^{i\theta_q}$. Then, in the chiral perturbation theory
language, we have the following terms from (\ref{IntL}) and
(\ref{instInt})
$$
-M\Lambda^3\cos\theta_\lambda-mv^3\cos\theta_q+\langle {\cal L}_{\rm
inst}\rangle .
$$
The problem is to calculate the expectation value of ${\cal L}_{\rm
inst}$. For this purpose, we insert the identity,
 $$
 {\bf 1}= |0\rangle\langle
0|+|\eta_\lambda\rangle\langle\eta_\lambda|
+|\eta_q\rangle\langle\eta_q|
+\cdots+(|\eta_\lambda\rangle\langle\eta_\lambda|)^N
|\eta_q\rangle\langle\eta_q|+\cdots
 $$
where momenta sum is understood. It can be schematically shown as
Fig. \ref{tHooft}. The uncondensed fermion lines are connected with
the current masses.

Thus, we obtain an effective interaction Hamiltonian of the form,
\begin{align}
{\cal
H}_{\chi}=&-M\Lambda^3\cos\theta_\lambda-mv^3\cos\theta_q\nonumber\\
&- f_0\Lambda^{3N}v^3\cos[N\theta_\lambda+\theta_q+(N+1)\theta]\nonumber\\
&-Nf_1 M\Lambda^{3(N-1)}v^3\cos[N\theta_\lambda+\theta_q+(N+1)\theta
+\alpha_\lambda]\nonumber\\
&-f_2 m\Lambda^{3N}\cos[N\theta_\lambda+\theta_q+(N+1)\theta
+\alpha_q]+\cdots\nonumber
\end{align}
where $f_0,f_1$ and $f_2$ are constants. Let us choose $M$ and $m$
real, i.e. take $\alpha_\lambda=\alpha_q=0$. For two condensates and
one axion, the $3\times 3$ mass matrix is
\begin{equation}
M^2=\left(\begin{array}{ccc}
\frac{(N^2\tilde\Lambda^4+M\Lambda^3)}{F_\lambda^2}, &
\frac{N\tilde\Lambda^4}{F_\lambda F_q},&
\frac{N(N+1)\tilde\Lambda^4}{F_\lambda F_a}\\
\frac{N\tilde\Lambda^4}{F_\lambda F_q}
&\frac{\tilde\Lambda^4+mv^3}{F_q^2}
 & \frac{(N+1)\tilde\Lambda^4}{F_qF_a}\\
\frac{N(N+1)\tilde\Lambda^4}{F_\lambda F_a}
&\frac{(N+1)\tilde\Lambda^4}{F_qF_a} &
\frac{(N+1)^2\tilde\Lambda^4}{F_a^2}
\end{array}\right)
\end{equation}
where
$$
\tilde\Lambda^4=f_0\Lambda^{3N}v^3N^2+f_1 M\Lambda^{3(N-1)}v^3N^3
+f_2 m\Lambda^{3N}N^2+\cdots
$$
where ellipses denote higher power terms of $m$ and $M$.  In the
limit of $\tilde\Lambda\gg M,m,$ we obtain three masses as
\begin{align}
\begin{split}
&\eta^{\prime 2}_\lambda=(a^2_\lambda+a^2_q+a^2_a)\tilde\Lambda^2\\
&\eta^{\prime 2}_q=\frac{(a^2_q+a^2_a)M\Lambda^3/F^2_\lambda
+(a^2_\lambda+a^2_a)mv^3/F_q^2}{(a^2_\lambda+a^2_q+a^2_a)}\\
&m_a^2=\frac{mv^3}{(1+\alpha_q)F_q^2
+(1+\alpha_\lambda)Z\beta F^2_\lambda}\\
\end{split}\label{masses}
\end{align}
where
\begin{align}
&Z=m/M,\ \alpha_q=a_q^2/a_a^2,\ \alpha_\lambda=a_\lambda^2/a_a^2,\
\beta=v^3/\Lambda^3,\nonumber\\
&a_\lambda=N\tilde\Lambda/F_\lambda,\ a_q=\tilde\Lambda/F_q,\
a_a=(N+1)\tilde\Lambda/F_a.
\end{align}
From (\ref{masses}), with a large  $m$ there is no pseudoscalar
which is almost massless. It is equivalent to saying that the QCD
instanton potential is lower than the hidden-sector instanton
potential.  Then the decay constant of the QCD axion is at the
string scale even if we have lowered some axion decay constant to a
lower scale.

\subsection{Squark condensation}

\begin{figure}[t]
\centering \epsfig{figure=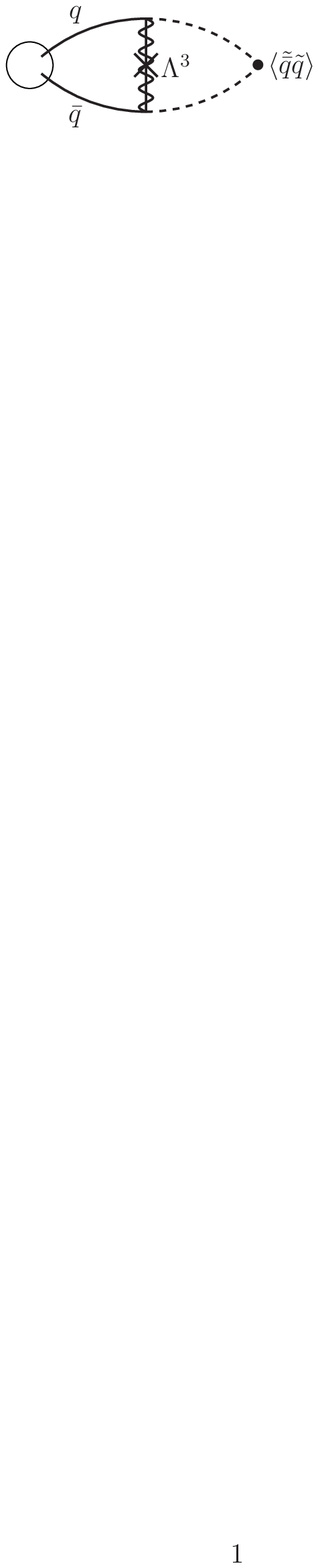, width=6cm,
bbllx=180,bblly=690, bburx=420,bbury=770 }
 \caption{A contraction of
the hidden-sector quark through one loop effect. Chiral symmetry is
broken by squark condensation.}\label{squarkcond}
\end{figure}

To reduce the hidden sector instanton potential drastically, we may
introduce a massless quark. Then from Eq. (\ref{masses}), the axion
mass is vanishing.  But the question is whether this masslessness of
the hidden-sector quark is supported or not in our scenario with
{\it supersymmetry}. The chiral symmetry is present  when we
introduced a massless hidden-sector quark. This chiral symmetry can
be broken by a squark condensation, not breaking supersymmetry. So
the hidden-sector quark is expected to obtain a mass of order
$\langle \tilde{\bar q}\tilde q\rangle/(8\pi^2)\Lambda$. In Fig.
\ref{tHooft}, the condensation of the hidden-sector quark cannot
appear but only the effective mass can be present as shown in Fig.
\ref{squarkcond}.

 Thus, with $n$
hidden-sector quarks, we expect the height of the hidden-sector
instanton potential is
\begin{equation}
{M}\Lambda^3\left(\frac{\langle \tilde{\bar q}\tilde q\rangle}{
8\pi^2\Lambda^2}\right)^n.
\end{equation}
Here we inserted one power of the hidden-sector gaugino mass.
Without the gaugino mass, i.e. without the gravitational
effect($M\sim\Lambda^3/M_P^2$), all insertions in the 't Hooft
determinental interaction involve condensation and cannot contribute
to the axion mass. It contributes to the $\eta^\prime_\lambda$ mass
in  Eq. (\ref{masses}). At this level, $\langle \tilde{\bar q}\tilde
q\rangle$ is not determined. There is no constraint on the scale of
the hidden-sector squark condensation which can be very small. If we
take  an arbitrary value of order $10^{-3}\Lambda$, the suppression
factor is of order $\sim 10^{-5n}$ which must bring down the
instanton potential below the QCD scale for $n>8$.

\section{Conclusion}

In this paper, we discussed two plausible mechanisms toward lowering
the decay constants of superstring axions. For the MD-axions winding
a nearly vanishing 2-cycle background in the warped region, the
localization is a plausible mechanism for lowering the axion decay
constant. Here, the Wen-Witten world-sheet instanton problem must be
evaded so that a large superpotential is not generated.  Two axions
are needed if we require the hidden sector confining force. In this
case, the axion mixing must be studied to pinpoint what is the decay
constant corresponding to the QCD axion. To lower the decay constant
of the QCD axion much below the string scale, it is pointed out that
it is required to make the hidden-sector instanton potential be
lowered below the QCD instanton potential. Even if we introduced
massless hidden-sector quarks, this problem is nontrivial because of
the chiral symmetry breaking via the hidden-sector squark
condensation. It was pointed out that it is possible with
sufficiently many hidden-sector quarks with a low value of the the
hidden-sector squark condensation.

\vskip 0.5cm \noindent [Note added]: During this work, a work on the
related topic appeared by P. Svrcek and E. Witten, hep-th/0605206.

\acknowledgments{We thank Kiwoon Choi for a useful communication.
This work is supported in part by the KRF ABRL Grant No.
R14-2003-012-01001-0. J.E.K. is also supported in part by the KRF
grants, Sundo Grant No. R02-2004-000-10149-0 and Star Grant No.
KRF-2005-084-C00001.}


\begin{thebibliography}{99}

\def\apj#1#2#3{Astrophys.\ J.\ {\bf #1} (#3) #2}
\def\ijmp#1#2#3{Int.\ J.\ Mod.\ Phys.\ {\bf #1} (#3) #2}
\def\mpl#1#2#3{Mod.\ Phys.\ Lett.\ {\bf A#1} (#3) #2 }
\def\nat#1#2#3{Nature\ {\bf #1} (#3) #2}
\def\npb#1#2#3{Nucl.\ Phys.\ {\bf B#1} (#3) #2}
\def\plb#1#2#3{Phys.\ Lett.\ {\bf B#1} (#3) #2}
\def\prd#1#2#3{Phys.\ Rev.\ {\bf D#1} (#3) #2}
\def\prl#1#2#3{Phys.\ Rev.\ Lett.\ {\bf #1} (#3) #2}
\def\prp#1#2#3{Phys.\ Rep.\ {\bf #1} (#3) #2}
\def\sjnp#1#2#3{Sov.\ J.\ Nucl.\ Phys.\ {\bf #1} (#3) #2}
\def\zp#1#2#3{Z.\ Phys.\ {\bf #1} (#3) #2}
\def\jhep#1#2#3{JHEP\ {\bf #1} (#3) #2}
\def\jcap#1#2#3{JCAP\ {\bf #1} (#3) #2}
\def\epjc#1#2#3{Euro. Phys. J.\ {\bf C#1} (#3) #2}
\def\rmp#1#2#3{Rev. Mod. Phys.\ {\bf #1} (#3) #2}


\bibitem{kimprp87} For a review, see, J. E. Kim, \prp{150}{1}{1987}.

\bibitem{peccei} R. D. Peccei and H. R. Quinn, \prl{38}{1440}{1977}.


\bibitem{Witten84} E. Witten, \plb{149}{351}{1984}.

\bibitem{Witten85} E. Witten, \plb{153}{243}{1985}.

\bibitem{kim85} K. Choi and J. E. Kim, \plb{154}{393}{1985}.

\bibitem{nilles03} J. E. Kim and H. P. Nilles,
\plb{553}{1}{2003}; J. E. Kim, H. P. Nilles, and M.
Peloso, \jcap{0501}{005}{2005}.

\bibitem{RSI} L. Randall and R. Sundrum, \prl{83}{3370}{1999}.

\bibitem{KS98} I. R. Klebanov and M. J. Strassler, \jhep{0008}{052}
{2000}.

\bibitem{GKP02} S. B. Giddings, S. Kachru and J. Polchinski,
\prd{66}{106006}{2002}.

\bibitem{Strominger86} A. Strominger,  \npb{274}{253}{1986}.

\bibitem{Becker} K. Becker, M. Becker, J.-X. Fu,
L.-S. Tseng, and S.-T. Yau,  hep-th/0604137, based on the original
mathematical work, J.-X. Fu and S.-T. Yau,  hep-th/0604063.

\bibitem{intst} C. P. Burgess, L. E. Iba\~nez, and F. Quevedo,
\plb{447}{257}{1999}.

\bibitem{choikim85} K. Choi and J. E. Kim, \plb{165}{71}{1985}.

\bibitem{kim99} J. E. Kim,
\jhep{9905}{022}{1999}; \jhep{0006}{016}{2000}.

\bibitem{choiM} K. Choi, \prd{56}{6588}{1996}; \prd{62}{043509}{2000}.

\bibitem{conlon} J.~P.~Conlon,
  hep-th/0602233.


\bibitem{CHSW} P. Candelas, G. Horowitz, A. Strominger, and E. Witten,
\npb{258}{46}{1985}.

\bibitem{GS84} M. Green and J. H. Schwarz, \plb{149}{117}{1984}.



\bibitem{hubsch} T. H\"ubsch,  Comm. Math. Phys. {\bf 108} (1987)
291; B. R. Greene, K. H. Kirklin, P. J. Miron, and G. G. Ross,
\npb{278}{667}{1986}.

\bibitem{Becker03} K. Becker, M. Becker, K. Dasgupta, and P. S. Green,
\jhep{0304}{007}{2003};
K. Becker, M. Becker, P. S. Green, K. Dasgupta, and E. Sharpe,
\npb{678}{19}{2004}.

\bibitem{WenWitten} X. G. Wen and E. Witten, \plb{166}{397}{1986}.



 \bibitem{tHooft} G. 't Hooft, \prd{14}{3432}{1976}
 and \prd{18}{2199}{1978}(E).

\bibitem{U1thooft} G. 't Hooft, \prp{142}{357}{1986}.


\end{thebibliography}
\end{document}